\documentclass[twoside]{article}
\usepackage{qic,epsfig}

\newcommand{\bra}[1]{\left\langle #1 \right|}
\newcommand{\ket}[1]{\left|#1\right\rangle}
\newcommand{\braket}[2]{\left\langle#1 |  #2\right\rangle}

\newcommand{\abs}[1]{\left|#1\right|}
\newcommand{\Tr}{\textrm{Tr}}

\begin{document}
\setlength{\textheight}{8.0truein}    

\runninghead{Quantumness, Generalized Spherical $2$-Design, and
Symmetric Informationally Complete POVM}
            {Isaac H. Kim}

\normalsize\textlineskip
\thispagestyle{empty}
\setcounter{page}{1}

\copyrightheading{0}{0}{2003}{000--000}

\vspace*{0.88truein}

\alphfootnote

\fpage{1}

\centerline{\bf
Quantumness, Generalized 2-Design} \vspace*{0.035truein}
\centerline{\bf and Symmetric Informationally Complete POVM}
\vspace*{0.37truein} \centerline{\footnotesize
Isaac H. Kim\footnote{email address : griffys@mit.edu }}
\vspace*{0.015truein} \centerline{\footnotesize\it Department of
Physics, Massachusetts Institute of Technology, 229 Vassar Street}
\baselineskip=10pt \centerline{\footnotesize\it Cambridge, MA 02139,
USA} \vspace*{10pt} \publisher{(received date)}{(revised date)}

\vspace*{0.21truein}

\abstracts{
C. A. Fuchs and M. Sasaki defined the quantumness of a set of
quantum states in \cite{Quantumness}, which is related to
the fidelity loss in transmission of the quantum states through a
classical channel. In \cite{Fuchs}, Fuchs showed that in
$d$-dimensional Hilbert space, minimum quantumness is
$\frac{2}{d+1}$, and this can be achieved by all rays in the space.
He left an open problem, asking whether fewer than $d^2$ states can
achieve this bound. Recently, in a different context, A. J. Scott
introduced a concept of generalized $t$-design in \cite{GenSphet},
which is a natural generalization of spherical $t$-design. In this
paper, we show that the lower bound on the quantumness can be
achieved if and only if the states form a generalized 2-design. As a
corollary, we show that this bound can be only achieved if the
number of states are larger or equal to $d^2$, answering the open
problem. Furthermore, we also show that the minimal set of such
ensemble is Symmetric Informationally Complete POVM(SIC-POVM). This
leads to an equivalence relation between SIC-POVM and minimal set of
ensemble achieving minimal quantumness. }{}{}

\vspace*{10pt}

 \vspace*{3pt}
\communicate{to be filled by the Editorial}

\vspace*{1pt}\textlineskip    
\section{Introduction}
\noindent
 One of the major discrepancy between classical physics
and quantum mechanics is the difference in the measurement
procedure. In classical physics, in principle, we can perform
arbitrarily precise measurement. However, in the realm of quantum
mechanics, this is prohibited by the laws of nature. For instance,
if we want to measure the momentum and the position of a particle,
we cannot measure them simultaneously with an indefinite accuracy
due to the following relation.
\begin{displaymath}
\Delta x \Delta p \geq \frac{\hbar}{2} \nonumber
\end{displaymath}

 Then, is there any good measure of `how quantum' a system is? There
is a wide belief - although it has not been explicitly proven - that
the notion of quantumness can be only defined on an ensemble of
states.\cite{Quantumness} For instance, assume we have a single
normalized state $\ket{\psi_1}$, and we want to measure `how
quantum' this state is. If we choose an appropriate set of
measurement operators(i.e. the POVM trivially constructed from the
projectors of the orthonormal set of basis containing
$\ket{\psi_1}$), we can measure the state with certainty. That is,
there is no quantum mechanical effect that differs from the
classical mechanics if we use the optimized measurement. However, if
we have an ensemble of states which has nonzero overlap between
them, the probability of making the wrong guess becomes strictly
larger than 0 even if we use the optimal strategy. Based on this
idea, the notion of quantumness was first proposed by C. A. Fuchs
and M. Sasaki \cite{Quantumness} in 2003 with several open problems.
In this paper, We will use the definition of the quantumness as the
one appearing in \cite{Quantumness}. Readers should be always aware
that the quantumness is an inverted measure. That is, higher
quantumness corresponds to higher classicality.

After his work with M. Sasaki in 2003, C. A. Fuchs showed that
symmetric informationally complete ensemble produces a minimal
quantumness\cite{Fuchs} and left an open problem which asked, ``Is
it possible to construct an ensemble that generates a minimal
quantumness with strictly less than $d^2$ elements?" In this paper,
We will give the answer to the question. It turns out that we need
at least $d^2$ ensemble of states to achieve the minimal
quantumness, where $d$ is the dimension of the Hilbert space.
Furthermore, the corresponding ensembles form SIC-POVM.

The remainder of the paper will have the following structure. In
Section 2, we will introduce several definitions that leads to a
definition of quantumness, with some preliminary results. This
section will be mostly based on \cite{Quantumness}. In Section 3, we
will have a brief introducion to frame theory and several new
definitions such as \emph{t-extended density operator} and
\emph{generalized $t$-design}. This section will be mainly based on
\cite{GenSphet} and \cite{SIC-POVM}, with some generalizations.
Section 4 will be the core of this paper. Here we will prove that
ensembles which are generalized $2$-design achieves minimal
quantumness and vice versa. Also, as a corollary, we will show that
minimal number of states to achieve minimal quantumness is $d^2$. In
section 5, we will show that generalized $2$-design with minimal
elements are SIC-ensemble, thereby showing the connection between
SIC-POVM and minimal ensembles achieving the minimal quantumness.
 Finally, conclusion with
remaining open problem will be discussed in Section 6. Throughout
the paper, the Hilbert space is assumed to be $d$ dimension unless
it is specified otherwise.

\section{Definitions and Preliminary Results}
\noindent
 Alice sends a quantum message $\mathcal{P}=\{\Pi_i, \pi_i
\}$ to Bob. $\Pi_i =\ket{\psi_i}\bra{\psi_i}$ where each
$\ket{\psi_i}$ is normalized. $\pi_i$ is the probability of
appearance for each quantum states. In between Alice and Bob, the
eavesdropper Eve intercepts the message. To do so, Eve has to
measure the states that is sent from Alice and must reconstruct it.
Eve's measurement is denoted as $\mathcal{E}=\{ E_b \}$, where $E_b$
forms a POVM. Her state reconstruction strategy is $\mathcal{M}$,
which is a mapping $b \to \sigma_b$. Here $\sigma_b$ is a density
matrix. The \emph{average fidelity} $F(\mathcal{E},\mathcal{M})$ is
the fidelity averaged over all possibilities.
\begin{eqnarray}
F_{\mathcal{P}}(\mathcal{E},\mathcal{M}) &=\sum_{b,i}p(b,i)\Tr(\Pi_i
\sigma_b)  \\ & \qquad = \sum_{b,i}\pi_i\Tr(\Pi_i E_b) \Tr(\Pi_i
\sigma_b)
\end{eqnarray}

The maximal fidelity that can be achieved by changing Eve's strategy
is defined as the \emph{accessible fidelity}.
\begin{equation}
F_{\mathcal{P}}=\sup_{\mathcal{E}}\sup_{\mathcal{M}}F_{\mathcal{P}}(\mathcal{E},
\mathcal{M})
\end{equation}
By changing the probabilities $\pi_i$, one can minimize the
\emph{accessible fidelity}. This is the quantumness of the state
ensemble $\{\Pi_i \}$.
\begin{equation}
Q_{\{\Pi_i \}}=\inf_{\mathcal{P}}F_{\mathcal{P}}
\end{equation}

 Since the fidelity can be made arbitrarily close to 1 for
orthogonal states, the maximum quantumness is $1$, which corresponds
to the classical case. C. A. Fuchs and M. Sasaki proved that the
smallest quantumness that can be achieved in $d$-dimensional Hilbert
space is $\frac{2}{d+1}$, and one example of such ensemble is the
unitarily invariant ensemble.\cite{Quantumness} In his later work,
he showed that there exists a discrete ensemble that achieves the
smallest quantumness with $d^2$ elements, which is a symmetric
informationlly complete ensemble(SIC-ensemble).\cite{Fuchs}

SIC-ensemble is a uniformly distributed ensemble of $d^2$ normalized
vectors in a $d$-dimensional Hilbert space $\mathcal{H}_d$,
satisfying the following condition.
\begin{eqnarray}
\Tr(\Pi_i \Pi_j) & = \frac{1}{d+1} &\textrm{for} \quad i\neq j \nonumber\\
& = 1  \quad &\textrm{for} \quad i=j
\end{eqnarray}
 The SIC-ensemble was coined from a
symmetric informationally complete POVM (SIC-POVM), and there are
many papers about this set of
states.\cite{SIC-POVM}\cite{Zauner}\cite{Appleby} One remarkable
fact is that the subnormalized projection operators of these
states($\frac{1}{d}\Pi_k$) form a POVM.\cite{Zauner} However, note
that the existence of such states in all dimensions has not been
proven yet, even though it is highly believed to be
so.\cite{SIC-POVM}

\section{Frame Theory and Generalized $t$-design}
\noindent Frame is a generalization of basis sets. For a
$d$-dimensional Hilbert space, a collection of vectors
$\ket{\psi_k}$ is a frame if there exist constants $0<a\leq b
<\infty$ such that
\begin{equation}
a\abs{\braket{\phi}{\phi}}^2 \leq
\sum_{k}\abs{\braket{\phi}{\psi_k}}^2 \leq
b\abs{\braket{\phi}{\phi}}^2
\end{equation}
for all $\ket{\phi}$ in the Hilbert space. $a$ and $b$ are called as
\emph{frame bounds}, and the frame is \emph{tight} if $a=b$. Now we
define the frame operator $S$.
\begin{equation}
S=\sum_{k}\ket{\psi_k}\bra{\psi_k}
\end{equation}
where $\ket{\psi_k}$ is a state that lives in the Hilbert space.

Generalized $t$-design is a natural generalization of spherical $t$-design,
in a sense that it allows the set of states to be uncountable and the
probability distribution to be nonuniform. Note that spherical $t$-design
is a set $\mathcal{D}_t=\{ \ket{\psi_k} \}$ of normalized vectors that satisfies
\begin{equation}
\Tr(\otimes_{j=1}^t A_j \frac{1}{\abs{\mathcal{D}_t}}(\ket{\psi_k}\bra{\psi_k})^{\otimes t}) = \Tr(\otimes_{j=1}^t A_j \int_{\mathcal{S}_d} d\psi (\ket{\psi} \bra{\psi})^{\otimes t}) \quad \forall A_j,
\end{equation}
where $\mathcal{S}_d$ is a  set of normalized states in the $d$-dimensional
Hilbert space.

Similarly, \emph{generalized $t$-design} can be defined as a set of
states $\mathcal{D}_t$ together with probability distribution $P$
over $\mathcal{D}_t$ such that $\int_{\mathcal{D}_t} dP =1$ that
satisfies the following.
\begin{equation}
\Tr(\otimes_{j=1}^t A_j \int_{\mathcal{D}_t} dP(\psi) (\ket{\psi}\bra{\psi})^{\otimes t}) = \Tr(\otimes_{j=1}^t A_j \int_{\mathcal{S}_d} d\psi (\ket{\psi} \bra{\psi})^{\otimes t}) \quad \forall A_j,
\end{equation}
Therefore, $(\mathcal{D}_t, P)$ is a generalized $t$-design if and only if
$\int_{D_t} dP(\psi) \ket{\psi} \bra{\psi} = \int_{\mathcal{S}_d} d\psi \ket{\psi} \bra{\psi}$.
In \cite{GenSphet}, the author uses a different definition of generalized
$t$-design, which is shown to be equivalent with the definition of this paper.
He shows this by proving
\begin{equation}
\int_{\mathcal{S}_d} d\psi \ket{\psi} \bra{\psi} = \frac{t!(d-1)!}{(t+d-1)!} \Pi_{sym},
\label{eq:t-DesignDefinitionConnection}
\end{equation}
where $\Pi_{sym}$ is a projector onto a  symmetric subspace in
$\mathcal{H}_d^{\otimes t}$. Note that  the lefthand side of
Eq.\ref{eq:t-DesignDefinitionConnection} is invariant under the
conjugation of $U^{\otimes t}$, where $U \in U(d)$. By Schur's lemma,
this must be a multiple of identity transformation on a symmetric subspace.
Imposing a trace condition on both sides, we arrive at Eq.\ref{eq:t-DesignDefinitionConnection}.

 Let us define a frame
operator with trace $1$ as a \emph{unit-trace frame operator}. From
any $S$, we can trivially construct $S'=\frac{S}{\Tr(S)}$ which is a
\emph{unit-trace frame operator}. Since this operator is positive,
one can see that there exists an equivalent density operator. From
this, we can generalize the theorem from Benedetto and
Fickus\cite{FrameBound}.
\begin{theorem}
Let $S$ be a unit-trace frame operator constructed from
$\{\ket{\psi_k}\}_{k=1}^{n}$, where each $\ket{\psi_k} \in
\mathcal{H}_d$. Then the following inequality holds.
\end{theorem}
\begin{equation}
\textrm{Tr}(S^2)\geq \max(\frac{1}{n},\frac{1}{d})
\end{equation}
\emph{The bound is achieved if and only if $\{\ket{\psi_k}\}$
consists of orthogonal vectors with uniform norm, when $n\leq d$, or
is a tight frame, when $n\geq d$.}

\textbf{Proof} Let us denote the eigenvalues of $S$ as $\lambda_k$
in a nonincreasing order. The number of eigenvalues are at most
$l=\min(n,d)$. Notice the following equation.
\begin{equation}
\Tr(S^2)=\sum_{k=1}^{n} \lambda_{k}^2
\end{equation}
Under the constraint $\sum_{k=1}^{n} \lambda_k =1$ and $\lambda_k
\geq 0$ for all $k$, the equality holds if and only if
$\lambda_k=\frac{1}{l}$ for all $k$. For $n\leq d$,
$S=\frac{\Pi_n}{n}$, where $\Pi_n$ is a projector onto a
$n$-dimensional subspace. Therefore, the vectors must be orthogonal
to each other and have uniform norm. For $n \geq d$,
$S=\frac{1}{d}I$, implying $\{\ket{\psi_k}\}$ is a tight frame. One
can also see that $dS=I$, which means
$\{d\ket{\psi_k}\bra{\psi_k}\}_{k=1}^{n}$ forms a POVM.

 Now we define a $t$\emph{-extended density operator}. Assume we have a
\emph{unit-trace frame operator} $S=\sum_{k} \pi_{k}
\ket{\psi_k}\bra{\psi_k}$ from ensemble $\{\Pi_i, \pi_i\}$. We
define the $t$\emph{-extended density operator} to be the following.
\begin{equation}
S_t=\sum_{k}\pi_k \ket{\psi_{k}}^{\otimes t} \bra{\psi_k}^{\otimes
t}
\end{equation}

From Theorem 1, using the equality condition, one can see that
$S_t$ with $\{\Pi_k, \pi_k \}_{k=1}^{n}$, $n\geq {}_{t+d-1}C_{d-1}$
is a generalized $t$-design if and only if
\begin{equation}
\Tr(S_t^2)= \frac{t!(d-1)!}{(t+d-1)!},
\end{equation}
and this is a global minimum of $\Tr(S_t^2)$. The sequence of
derivation we have shown so far resembles that of a \emph{spherical
t-design} in \cite{SIC-POVM}. One major difference is that in
\cite{SIC-POVM}, the probabilities corresponding to the states were
uniform while here the condition is relaxed.

\section{Generalized $2$-Design and Ensembles Achieving Minimal Quantumness}
\noindent Let us start with the definition of average fidelity.
\begin{equation}
F_{\mathcal{P}}(\mathcal{E},\mathcal{M}) = \sum_{b,i}\pi_i\Tr(\Pi_i
E_b) \Tr(\Pi_i \sigma_b)\nonumber
\end{equation}
 Let $\tilde{\rho}$ be the $2$-extended density operator
$\sum_{i}\pi_i \Pi_i^{\otimes 2}$ and let $X\equiv \sum_{b} E_b
\otimes \sigma_b$. We can interprete this as $E_b$ living on a
\emph{measurement operator space} and $\sigma_b$ living on a
\emph{reconstruction operator space}. One can see that the average
fidelity can be simply rewritten in the following expression.
\begin{equation}
F_{\mathcal{P}}(\mathcal{E},\mathcal{M})=\Tr(\tilde{\rho} X)
\end{equation}

POVM elements $E_b$ can be expressed as a projection of
subnormalized vectors, and the state-reconstruction density operator
$\sigma_b$ can be expressed as a linear sum of normalized projection
operators $\sum_i p_i \ket{\phi_i}\bra{\phi_i}$. Consider an
ensemble which forms a generalized  2-design. We use the state
reconstruction strategy $\{E_b=\lambda_b
\ket{\phi_b}\bra{\phi_b}\textrm{ } | \sum_b E_b = I, 0< \lambda_b
\leq 1\}$ , $\{ \sigma_b=\sum_k p_k \ket{\varphi_k^b}\bra{
\varphi_k^b}  \textrm{ }| \sum_k p_k=1\}$, which captures any
possibility. Therefore, the corresponding average fidelity becomes
the following.
\begin{equation}
\int d\psi \sum_{b,k} p_k \lambda_b \abs{\braket{\phi_b}{\psi}}^2
\abs{\braket{\varphi_k^b}{\psi}}^2 \label{eq:SpecialAF}
\end{equation}

\begin{lemma}
Ensembles which form generalized $2$-design achieves minimal 
quantumness. \label{lemma:1}
\end{lemma}
\textbf{Proof} Notice the following Cauchy-Schwarz inequality for nonnegative
functions $f$ and $g$ over a measure $\Omega$.
\begin{equation}
\int d\Omega f(\Omega) g(\Omega) \leq \sqrt{\int d\Omega f(\Omega)^2
\int d\Omega g(\Omega)^2}
\end{equation}
We can directly apply this inequality to Eq.\ref{eq:SpecialAF}.
Since the equality holds if and only if
$\abs{\braket{\phi_b}{\psi}}^2= \abs{\braket{\varphi_k^b}{\psi}}^2$
for all $\ket{\psi}$, $\ket{\phi_b}$ and $\ket{\varphi_k^b}$ must be
same up to a phase for all $b$ and $k$. When this equality holds, we
obtain an expression for the average fidelity,
\begin{eqnarray}
F_{\mathcal{P}}(\mathcal{E},\mathcal{M})& = d\int d\psi \abs{\braket{\psi_i}{\psi}}^4 \label{eq:InfInt}\\
&= \frac{2}{d+1} \label{eq:Result}
\end{eqnarray}
where $\ket{\psi_i}$ is a dummy state for integration. Therefore,
any ensemble which is a generalized $2$-design
achieves minimal quantumness. Furthermore, the optimal strategy is
to measure the states with rank-1 POVM and reconstruct the
corresponding states. This agrees with the result from \cite{Fuchs}.

Assume the converse of this statement is incorrect.
This means the \emph{$2$-extended density operator} generated by the
ensemble is not identical to the projector onto the symmetric
subspace. This quantity can be represented as the following
equation.
\begin{equation}
\Delta \tilde{\rho} = \tilde{\rho} - \frac{2}{d(d+1)}\Pi_{sym}
\end{equation}

Note that by the construction of $\Delta \tilde{\rho}$, it is
traceless. Similarly, we may reexpress $\tilde{X}(\tilde{\rho})$ as
$\frac{2}{d+1}\Pi_{sym}+ \tilde{X}'(\tilde{\rho})$. In this case, we
have a stronger relation then $\Tr(\tilde{X}')=0$. Since $\Tr_2(X)
\equiv \sum_b E_b \Tr(\sigma_b)=I$ for any $X$, we get $\Tr_2(\tilde{X}')=0$.
Now we can express the accessible fidelity in a following form,
owing to the fact that $\Delta \tilde{\rho}$ and $\tilde{X}'$ are
traceless.
\begin{equation}
F_{\mathcal{P}}=\frac{2}{d+1}+\Tr(\Delta \tilde{\rho}
\tilde{X}'(\tilde{\rho}))
\end{equation}

\begin{lemma}
If $\Delta \tilde{\rho}\neq 0$, there exists $\tilde{X}'$ such that
$\Tr(\Delta \tilde{\rho} \tilde{X}'(\tilde{\rho}))\neq 0$. \label{lemma:2}
\end{lemma}
\textbf{Proof} Since $\Delta \tilde{\rho}$ is hermitian, we can
diagonalize it respect to some orthogonal basis. Since the trace is
invariant under unitary transform of the operators, we work in this
basis for $\tilde{X}'$ as well. Since we are working in the
symmetric subspace, we denote each of the diagonal entries as
$(i,j)$ where $(i,j)=(j,i)$. Suppose each diagonal entries of
$\Delta \tilde{\rho}$ are denoted as $\alpha_{(i,j)}$. The following
relation must hold, since $\Tr(\Delta \tilde{\rho})=0$.
\begin{equation}
\sum_{i=1}^d \sum_{j=1}^i\alpha_{(i,j)}=0
\end{equation}
Let us denote the diagonal entries of $\tilde{X}'$ as
$\beta_{(i,j)}$. Due to the constraint $\Tr_2(\tilde{X}')=0$, we
obtain a system of linear constraints.
\begin{equation}
\beta_{(i,i)}+\frac{1}{2}\sum_{j=1, j\neq i}^{d} \beta_{(i,j)}=0
\quad \forall i
\end{equation}
Using this equation, we can deduce the way to set $\Tr(\Delta
\tilde{\rho} \tilde{X}'(\tilde{\rho}))$ to $0$.
\begin{eqnarray}
\Tr(\Delta \tilde{\rho} \tilde{X}'(\tilde{\rho})) &= \sum_{(i,j)} \alpha_{(i,j)} \beta_{(i,j)} \\
&= \sum_{i=1}^d
\sum_{j=1}^{i-1} \alpha_{(i,j)} \beta_{(i,j)} + \sum_{i=1}^d
\alpha_{(i,i)} \beta_{(i,i)}
\\
&=\frac{1}{2}\sum_{i=1}^d \sum_{j=1, j\neq i}^{d} (\alpha_{(i,j)} -
\alpha_{(i,i)})\beta_{(i,j)}=0
\end{eqnarray}
Note that $\sum_{(i,j)}=\sum_{i=1}^d \sum_{j=1}^i$. Since
$\beta_{(i,j)}$ can have nontrivial values, to set $\Tr(\Delta
\tilde{\rho} \tilde{X}'(\tilde{\rho}))$ to zero regardless of the
choice of $\beta_{(i,j)}$, $\alpha_{(i,j)}$ must be a constant
respect to the indices $i$ and $j$. Since $\sum \alpha_{(i,j)}=0$ we
can conclude that $\alpha_{(i,j)}=0$ for all $i$ and $j$. Therefore,
unless $\Delta \tilde{\rho}$ is not equal to the null operator, we
can always find an operator which satisfies $\Tr(\Delta \tilde{\rho}
\tilde{X}'(\tilde{\rho}))\neq 0$. This ends the proof.

From Lemma \ref{lemma:1} and Lemma \ref{lemma:2}, 
we can conclude the following.
\begin{theorem}
Ensemble $\mathcal{P}(\Pi, \pi)$ is a generalized $2$-design if and
only if it achieves the minimal  quantumness.
\end{theorem}
\textbf{Proof} In Lemma \ref{lemma:1}, we already proved that
ensembles which are generalized $2$-design achieve minimal quantumness.
Therefore, the remaining part of the proof is the converse of that
statement. Note that we can prevent $X'$ from violating the
positivity of the operator $\tilde{X}$, by dividing $\tilde{X}'$
with a sufficiently large number. If it gives negative result, we
can simply change its sign, keeping the entries small enough to
prevent the violation of positivity. This means that if the ensemble
does not form a generalized $2$-design, we can always \emph{do
better} than the cases corresponding to the minimal quantumness.
Therefore, all ensembles which achieve minimal quantumness are
necessarily generalized $2$-design. 

Furthermore, minimal number of states needed for generalized
$2$-design is $d^2$, and the logic follows very closely that of
\cite{SIC-POVM}. In \cite{SIC-POVM}, the authors show that the
minimal number of states needed for spherical $2$-design is $d^2$ by
proving the following relation.
\begin{equation}
\int_{\mathcal{S}_d} d\psi \ket{\psi} \bra{\psi} A \ket{\psi}
\bra{\psi} = \frac{1}{d(d+1)} (A+ \Tr(A)I) \quad \forall A \in
\mathcal{B}(\mathcal{H}_d), \label{eq:RenEq}
\end{equation}
where $\mathcal{B}(\mathcal{H}_d)$ is a set of linear operators
on $\mathcal{H}_d$.

Note that the left hand side of Eq.\ref{eq:RenEq} is identical to
the action of generalized $2$-design on an element of
$\mathcal{B}(\mathcal{H}_d)$. Since the right hand side is clearly a
rank-$d^2$ operator, it follows that the minimal number of states
needed is $d^2$. For a detailed proof, we recommend to consult
section 2 of \cite{SIC-POVM}.

\section{Generalized $2$-design with $d^2$ elements are Spherical $2$-design}
\noindent Although spherical $2$-design is a proper subset of
generalized $2$-design, these two sets coincide if we only take care
of the minimal sets. Since generalized $2$-design is automatically a
generalized $1$-design as well, by using Lemma 1, we may find
following two formulae.
\begin{eqnarray}
&\Tr(\tilde{\rho}^2)=\frac{2}{d(d+1)}\label{eq:gs2design} \\
&\Tr(\rho^2)=\frac{1}{d}\label{eq:gs1design}
\end{eqnarray}
Setting $\lambda_{jk}\equiv \abs{\braket{\psi_j}{\psi_k}}^2$ for
$j\neq k$, we can reformulate Eq.\ref{eq:gs2design} and
Eq.\ref{eq:gs1design}.
\begin{eqnarray}
 & \sum_{j\neq k}p_j p_k \lambda_{jk}^2 + \sum_k
p_k^2=\frac{2}{d(d+1)}\label{eq:hyperellipsoid}\\
& \sum_{j\neq k}p_j p_k \lambda_{jk} + \sum_k p_k^2 = \frac{1}{d}
\label{eq:hyperplane}
\end{eqnarray}
Now let us find the range that the quadratic form $\sum_{j\neq k}
p_jp_k \lambda_{jk}^2$ can have under the constraint of
Eq.\ref{eq:hyperplane}. By using Lagrange's multiplier method, the
minimum can be found at points $\lambda_{jk}=\frac{1}{1-\alpha}
(\frac{1}{d}-\alpha)$, where $\alpha \equiv \sum_k p_k^2$. If the
number of ensembles is $d^2$, a bound on $\alpha$ becomes
$\frac{1}{d^2}\leq \alpha \leq 1$, where the former bound is
achieved if and only if $p_j=\frac{1}{d^2}$ for all $j$. From this
inequality, we can derive the following result.
\begin{equation}
\sum_{j\neq k}p_j p_k \lambda_{jk}^2 + \sum_k p_k^2 \geq
\frac{2}{d(d+1)}
\end{equation}
The bound is achieved if and only if $\alpha=\frac{1}{d^2}$,
$\lambda_{jk}=\frac{1}{d+1}$. Therefore, if a generalized $2$-design
exists for $d^2$ elements, it must be a spherical $2$-design as
well. Note that this argument cannot be applied to the number of
ensembles not equal to $d^2$.

\section{Discussion}
\noindent
 In this paper, we have shown that ensembles achieveing
minimal quantumness are generalized 2-design and vice versa. As a
corollary, we also showed that the minimal number of states needed
to achieve minimal quantumness is $d^2$. Furthermore, for these
ensembles, their optimal strategy is to measure states and simply
reconstruct the corresponding states, as shown in \cite{Fuchs}.
However, all the ensembles that have been known so far to achieve
the minimal quantumness are spherical $2$-design. These are the
unitarily invariant ensemble, SIC-ensemble, and ensemble formed by a
uniform distribution of mutually unbiased basis(MUB).
\cite{Quantumness}\cite{SIC-POVM}\cite{Fuchs}\cite{MUB} Although it
is possible to construct a generalized $2$-design which is not a
spherical $2$-design by merging two different spherical $2$-designs,
it will be interesting to see ensembles which achieve a minimal
quantumness and cannot be decomposed in such a trivial way. Whether
such examples exist or not is left as an open problem.

 Also, by showing the equivalence between the generalized
$2$-design and spherical $2$-design for $d^2$ ensembles, we found an
equivalent definition of the SIC-POVM in a totally different point
of view. This means we have a new way to prove the existence of the
SIC-POVM in all dimensions that has been unseen among the previous
literatures. The existence of SIC-POVM will follow if the minimal
quantumness can be achieved with only $d^2$ vectors. The usefulness
of this approach has not been investigated much at this point, but
we are hoping to see positive results.

\nonumsection{Acknowledgements} \noindent Author would like to thank Peter
Shor for his helpful discussions in writing this paper.

\nonumsection{References} 
\end{document}